\documentclass[12pt]{iopart}
\usepackage{iopams}
\usepackage{graphicx}
\usepackage{amssymb}
\usepackage{soul}
\usepackage{cancel}
\usepackage{color}

\begin{document}

\title[]{Modified fluctuation-dissipation theorem near non-equilibrium states and applications to the Glauber-Ising chain}

\newcommand{\bea}{\begin{eqnarray}}
\newcommand{\eea}{\end{eqnarray}}
\newcommand{\beq}{\begin{equation}}
\newcommand{\eeq}{\end{equation}}
\newcommand{\bit}{\begin{itemize}}
\newcommand{\eit}{\end{itemize}}
\newcommand{\Y}{\mathcal{Y}}
\newcommand{\ra}{\rightarrow}
\newcommand{\s}{\sigma}
\newcommand{\cf}{\{ \s \}}
\newcommand{\cfp}{\{ \s' \}}

\newcommand{\graph}[3]{
\begin{figure}[h!]
\includegraphics[scale=0.8]{#1}
\caption{#2}
\label{#3}
\end {figure}
}

\author{Gatien Verley$^1$, Rapha\"{e}l Ch\'{e}trite$^2$ and David Lacoste$^1$}
\address{$^1$ Laboratoire de Physico-Chimie Th\'eorique - UMR CNRS Gulliver 7083,
ESPCI, 10 rue Vauquelin, F-75231 Paris, France}
\address{$^2$ Laboratoire J. A. Dieudonn\'e, UMR CNRS 6621, Universit\'e de Nice Sophia-Antipolis, Parc Valrose, 06108 Nice Cedex 02, France}
\ead{gatien.verley@espci.fr}

\begin{abstract}
In this paper, we present a general derivation of a modified fluctuation-dissipation theorem (MFDT) valid near an arbitrary non-stationary state for a system obeying markovian dynamics. We show that the method to derive modified fluctuation-dissipation theorems near non-equilibrium stationary states used by J. Prost et al., PRL 103, 090601 (2009), is generalizable to non-stationary states. This result follows from both standard linear response theory and from a transient fluctuation theorem, analogous to the Hatano-Sasa relation. We show that this modified fluctuation-dissipation theorem can be interpreted at the trajectory level using the notion of stochastic trajectory entropy, in a way which is similar to what has been done recently in the case of MFDT near non-equilibrium steady states (NESS). We illustrate this framework with two solvable examples: the first example corresponds to a brownian particle in an harmonic trap submitted to a quench of temperature and to a time-dependent stiffness. The second example is a classic model of coarsening systems, namely the 1D Ising model with Glauber dynamics.

\end{abstract}

\maketitle

\section{Introduction}

It is a general rule that as a system gets smaller its fluctuations increase.
As a consequence, in small systems (like a colloidal particle or a biomolecule), thermodynamic quantities like work \cite{Jarzynski1997_vol78,Sekimoto1998_vol130} or heat
are only defined in a statistical sense. Exact relations between
the statistical distributions of these thermodynamic quantities, known as fluctuations relations, have been obtained about a decade ago. Such ideas have lead to the emergence of a
new field, concerned by the specificity of thermodynamics for small systems and which has been called stochastic thermodynamics.

Fluctuation relations hold very generally for a
large class of systems and arbitrarily far from equilibrium
\cite{Jarzynski1997_vol78,Gallavotti1995_vol74,Kurchan1998_vol31,Lebowitz1999_vol95,Crooks2000_vol61,Hatano2001_vol86}.
They provide fresh ideas for revisiting an old but central issue of statistical physics, namely the origin of irreversibility of macroscopic systems.
Furthermore, within the linear regime, these fluctuations relations lead to various new modified fluctuation-dissipation theorems (MFDT), which are interesting and valuable extensions of the classical fluctuation-dissipation theorem \cite{Callen1951_vol83,Kubo1966_vol29,Risken1989_vol,Marconi2008_vol461}.

Recently, three main routes have emerged to construct such generalizations of
the fluctuation-dissipation theorem:
\begin{itemize}
  \item In the first route opened by L.F. Cugliandolo et al. \cite{Cugliandolo1994_vol4} and  continued by E. Lippiello et al. \cite{Lippiello2005_vol71} and G. Diezemann \cite{Diezemann2005_vol72}, the response function is written as a sum of a time derivative of the correlation function (similar to the equilibrium FDT), plus an additive function, called the asymmetry,
 which vanishes under equilibrium conditions. A physical interpretation for this asymmetry has been missing for many years, until recently, M. Baiesi et al. \cite{Baiesi2009_vol103} propose to interpret it using a new concept called frenesy. This frenesy contains the time symmetric part of the non-equilibrium fluctuations.
  \item In the second route opened by T. Speck and U. Seifert \cite{Speck2006_vol74}, the modifications of the fluctuation-dissipation theorem can be related to the so-called local velocity, which originates in the local currents present in the non-equilibrium situation. This route was further extended and generalized by R. Ch\'etrite et al. \cite{Chetrite2008_vol} who also provided the MFDT with a Lagrangian frame interpretation  \cite{Chetrite2009_vol137}. These ideas have then been confirmed experimentally using colloidal particles confined to circular trajectories \cite{Gomez-Solano2009_vol103,Blickle2007_vol98}. In the end, it appears that the first and second routes are closely related and can be unified through the introduction of stochastic derivatives \cite{Chetrite2011_vol143}.
\item  In the third approach developed by J. Prost et al. \cite{Prost2009_vol103} (see also R. Ch\'etrite et al.  \cite{Chetrite2008_vol}), the modified fluctuation theorem valid  near non-equilibrium steady states (NESS) takes the standard equilibrium form except that it involves a new observable, function of the non-equilibrium steady state. This new observable is the system stochastic entropy, which here must be evaluated in the NESS \cite{Verley2011_vol93,Seifert2010_vol89}.
\end{itemize}

Surprisingly, while the first and second approach provide ways to construct generalizations of the fluctuation-dissipation for cases where the system is initially in an arbitrary non-equilibrium state, the third approach seems limited to systems close to a non-equilibrium stationary state.
One objective of the present paper is to show that the third approach can too be extended to systems close to a general non-stationary state, in a way which is closely related to the first approach. In view of this,  it appears that the three approaches provide closely related formulations to generalize the fluctuation-dissipation theorem to non-equilibrium situations.

Such generalizations could potentially lead to a broad range of applications to non-stationary or driven systems, such as glasses, spin glasses, coarsening systems, granular media and dense colloidal systems, for which violations of the fluctuation-dissipation theorem have been an active topic for many years \cite{Cugliandolo1994_vol4,Lippiello2005_vol71,Diezemann2005_vol72,Crisanti2003_vol36}.
Another more recent but promising field of applications of these ideas concerns biological systems \cite{Martin2001_vol98,Mizuno2007_vol315,Toyabe2010_vol104},
where the application of MFDT could possibly lead to new methods to probe these complex media.


In the next section, we show how to derive a MFDT classically, first using linear response theory, and then using more recent methods based on fluctuation relations. We then discuss an interpretation of the MFDT within stochastic thermodynamics, according to which the MFDT can be formulated in terms of a particular form of stochastic entropy. In the third section, we illustrate our framework with two pedagogical examples. In the first example, a Brownian particle placed in an harmonic potential is submitted to a quench of temperature and to a time-dependent stiffness, and in the second exemple, a 1D Ising chain obeying Glauber dynamics is submitted to a quench of temperature and then probed with a magnetic field.

\section{Modified Fluctuation-dissipation theorem (MFDT) for general non-stationary states}

\subsection{Stochastic modelling and definitions}
\label{definitions}
In the following, we derive a modified fluctuation-dissipation theorem (MFDT) for a system which is initially (at time $0$) in a general non-stationary state. The evolution of the system at all times is assumed to follow a continuous-time Markovian dynamics of a pure jump type \cite{Feller1940_vol48}. The transition rate to jump from a configuration $c$ to a configuration $c'$ is denoted $w_t(c,c')$, where the subscript $t$ indicates that we allow for time-dependent rates. We denote $\rho_t(c)$ the probability to be in state $c$ at time $t$. This quantity obeys the unperturbed master equation,
\beq
\frac{d \rho_t(c)}{dt} = \sum_{c'} [ \rho_t(c') w_t(c',c)  - \rho_t(c) w_t(c,c')  ] \,
\eeq
which can be written equivalently
\beq
\frac{d \rho_t(c)}{dt}  =  \sum_{c'} \rho_t(c') L_t(c',c), \label{eq evolution rho}
\eeq
in terms of the time-dependent markovian generator $L_t(c',c)$ defined by
\beq
L_t(c',c) = w_t(c',c) - \delta(c,c')\sum_{c''} w_t(c',c''). \label{def L}
\eeq
At time $t=0$, an arbitrary but given time-dependent perturbation $h_{t}$ is applied to the system, and we denote by $P_t(c, [h_t])$ the probability to observe the system in the state $c$ at a time $t$ in the presence of this perturbation. The notation $[h_t]$ emphasizes that the dependence is functional with respect to the perturbation. The evolution of the system for $t>0$ is controlled by the generator $L_t^{ h_t}$, which is defined similarly as in Eq.~\ref{def L} provided the rates $w_t(c,c')$ are replaced by perturbed rates $w^{h_t}_t(c,c')$. This generator can be expanded to first order in $[h_t]$
\beq
L_t^{h_t}=L_t+h_t N_t.
\eeq
In the following, it does matter whether the unperturbed dynamics is autonomous or not. If this dynamics is non-autonomous, {\it i.e.} proceeds from the application of a protocol, we still denote the generator by $L_t$ without specifying this protocol explicitly. In this case, the application of the perturbation $h_t$ for $t>0$ can be seen as an additional protocol.

Let us introduce $\pi_t(c,h)$ as the probability to observe the system in the state $c$ at a time $t>0$ in the presence of a constant (time independent) perturbation $h$, which obeys
\beq
\left( \frac{\partial \pi_t}{\partial t}  \right) (c,h) = \sum_{c'}  \pi_t(c',h) L_t^{h}(c',c).
\label{defpi}
\eeq

A key object for the following discussion is $\pi_t(c,h_t)$, which is constructed from $\pi_t(c,h)$ by replacing the time independent constant $h$ by the value of the perturbation at time $t$, namely $h_t=h(t)$. In the particular case where the perturbed dynamics with a constant $h$ is time independent (i.e. $L_t^{h} \equiv L^h$), the subscript $t$ in $\pi_t(c,h)$ may be dropped, and $\pi(c,h_t)$ becomes the "accompanying" distribution introduced in Ref.~\cite{Hanggi1982_vol88}.
We emphasize that $\pi_t(c,h_t)$ depends only on the perturbation at time $t$ unlike $P_t(c, [h_t])$ which depends functionally on the protocol history of the perturbation.
The dynamics of $\pi_t(c,h_t)$ is given by
\bea
 \frac{d}{dt} \left(\pi_t(c,h_t)\right) &=& \sum_{c'}  \pi_t(c',h_t) L_t^{h_t}(c',c) + \dot h_t \frac{\partial \pi_t(c,h_t)}{\partial h_t } \\
 &=& \sum_{c'} \pi_t(c',h_t) \left[ L_t^{h_t}(c',c) - \delta(c',c)\dot h_t \partial_{h_t} \psi_t(c,h_t)  \right], \label{pi_evo2}
\eea
where in the second equation one has introduced $\psi_t(c,h_t)=- \ln \pi_t(c,h_t)$ and an initial condition  $\rho_0(c)=P_0(c,0)=\pi_0(c,0)$ has been assumed.

The response function associated with a dynamic observable $A_{t}(c_t,h_t)$ reads, for a perturbation applied at an earlier time $t'>0$:
\beq
R(t,t')= \left. \frac{\delta \langle A_t(c_t,h_t) \rangle_{[h_t]}}{\delta h_{t'}} \right|_{h \rightarrow 0}.
\eeq
where $\langle .. \rangle_{[h_t]}$ represents an average with respect to the perturbed dynamics, and $\delta/\delta h_{t'}$ is our notation for functional derivatives.
We derive below a general formulation of a modified dissipation theorem for non-stationary states, which reads for $t>t'>0$
\beq
R(t,t')=-\frac{d}{dt'} \left\langle \left. \partial_h \psi_{t'}(c_{t'},h) \right|_{h \rightarrow 0}  A_{t}(c_{t},h_t) \right\rangle. \label{reponse}
\eeq
This relation qualifies for a modified fluctuation-dissipation because the response function is now expressed in terms of a correlation function of observables with respect to the
unperturbed dynamics, denoted $\langle .. \rangle$ \cite{Risken1989_vol}.
Note that when $A_t(c_t,h_t)=\partial_{h_{t}} \psi_{t}(c_{t},h_{t})$, the  modified fluctuation-dissipation theorem takes a more symmetric form, derived in Ref.~\cite{Prost2009_vol103}, under some specific conditions: these conditions are that the  initial state must be stationary and that the dynamics followed by the system at constant perturbation $h$ must be time independent ($L_t^{h} \equiv L^h$). Thus, as announced in the introduction, the present derivation extends the result developed in this reference to a more general initial condition and a more general dynamics.

\subsection{Derivation of MFDT from linear response theory}

This section shows how the modified fluctuation dissipation theorem of Eq.~\ref{reponse} follows from standard linear response theory. Starting from the master equation given above, one can generate a Dyson-type equation for the perturbed propagator $P(ct|c't')$, which is a fundamental result of linear response theory \cite{Kubo1966_vol29,Risken1989_vol,Hanggi1982_vol88}.
This propagator $P(ct|c't')$ represents the probability to find the system in the state $c$ at time $t$ given that it was in the state $c'$ at time $t'$ according to the perturbed dynamics, while $\rho(ct|c't')$ denotes the corresponding propagator for the unperturbed dynamics. When taken to first order in $[h_t]$, the Dyson equation \cite{Joachain1975_vol} for the propagator reads:
\begin{equation}
P(ct|c_0 0)= \rho(ct|c_0 0) + \int_0^t dt' h_{t'} \sum_{c'',c'} \rho(ct|c't') N_{t'}(c'',c') \rho(c'' t'|c_0 0).
\end{equation}
We then multiply this equation by an arbitrary observable, which we denote here $A_t(c)$ as a shorthand notation for $A_t(c,h_t)$. After integrating over the initial distribution $\rho_0(c_0)$, one obtains
\beq
\langle A_t(c_t) \rangle_{[h_t]} =  \langle A_t(c_t) \rangle + \int_0^t dt' h_{t'} \sum_{c,c'} A_t(c) \rho(ct,c't') B_{t'}(c')
\eeq
where we used the notation $\rho(ct,c't')=\rho(ct |c't') \rho_{t'}(c')$, and $B$ is the operator such that
\begin{equation}
B_{t'}(c') = \rho_{t'}(c')^{-1}  \sum_{c''} N_{t'}(c'',c') \rho_{t'}(c'').
\end{equation}
The main point of introducing $B$ is that it allows to write the response function in terms of a correlation function of two observables with respect to the unperturbed dynamics \cite{Risken1989_vol}:
\begin{equation}
R(t,t')= \langle B_{t'}(c_{t'})A_t(c_t)  \rangle.
\label{B form of MFDT}
\end{equation}
We can now use an expansion of the distribution $\pi_t$ to first order in $h$, $\pi_t(c,h)=\pi^{(0)}_t(c)+ h \pi^{(1)}_t(c)$. It follows from the master equation that the zeroth order solution is $\pi^{(0)}_t(c)=\rho_t(c)$, while the first order solution is
\beq
 \frac{\partial \pi_t^{(1)}(c)}{\partial t} = \sum_{c'}  \pi_t^{(0)}(c') N_t(c',c)+\pi_t^{(1)}(c') L_t(c',c). \label{pi order 1}
\eeq
Thus, the observable $B$ defined above can be written
\bea
B_t(c) &=& \pi_t^{(0)}(c)^{-1} \sum_{c'} N_t(c',c) \pi_t^{(0)}(c'), \\
       &=& \pi_t^{(0)}(c)^{-1} \left( \frac{\partial}{\partial t}  \pi_t^{(1)}(c)  - \sum_{c'}  L_t(c',c) \pi_t^{(1)}(c') \right ).
\eea
After substituting this in Eq.~\ref{B form of MFDT}, one obtains
\beq
R(t,t') = \sum_{c,c'} A_t(c) \rho(ct|c't')  \left( \frac{\partial}{\partial t'}  \pi_{t'}^{(1)}(c')  - \sum_{c''}  L_{t'}(c'',c') \pi_{t'}^{(1)}(c'') \right ).
\eeq
This form can be further transformed using the property that the unperturbed propagator $\rho(ct|c't')$ satisfies the backward Kolmogorov equation \cite{Gardiner1994_vol}
\beq
\partial_{t'} \rho(ct|c''t') = - \sum_{c'} L_{t'}(c'',c') \rho(ct|c't'),
\eeq
so that in the end
\beq
R(t,t') = \frac{d}{dt'} \left ( \sum_{c,c''}A_t(c)  \rho(ct|c''t') \pi_{t'}^{(1)}(c'') \right ),
\label{resp}
\eeq
which leads to Eq.~\ref{reponse} after using the relation
$\pi_{t'}^{(1)}(c) / \pi_{t'}^{(0)}(c) = \partial_h \ln \pi_{t'}(c,h)|_{h \ra 0}$.

\subsection{Derivation of MFDT from fluctuation relations}

For each path trajectory, we introduce the following functional
\beq
\Y_t=\int_0^t \dot h_\tau \partial_{h_\tau} \psi_\tau(c_\tau,h_\tau)  d\tau. \label{def Y}
\eeq
This functional $\Y_t$ has already appeared in Refs.~\cite{Chetrite2011_vol143,Chetrite2009_vol80} but in a different form. In the appendix A, we explain the connections between the different formulations.
The advantage of writing $\Y_t$ in the form of Eq.~\ref{def Y}, besides its simplicity, is that the similarity with the functionals introduced by Jarzynski \cite{Jarzynski1997_vol78} and Hatano-Sasa \cite{Hatano2001_vol86} is then very apparent.

In the same spirit as in the seminal works of Jarzynski and Hatano-Sasa, we consider below averages over trajectories with a weight $\Y_t$. To perform such averages, we
introduce the joint probability to be in the configuration $c$ at time $t$
with a value $\Y$ for the quantity $\Y_t$, $P_t(c,\Y)$, which is defined by
\beq
P_t(c,\Y) = \langle \delta(c-c_t) \delta(\Y-\Y_t) \rangle_{[h_t]}.
\eeq
This quantity obeys the following master equation
\beq
\frac{\partial P_t(c,\Y)}{\partial t} = \sum_{c'}  P_t(c',\Y) L_t^{ h_t}(c',c) -\dot h_t \frac{\partial \psi_t(c,h_t)}{\partial h_t} \frac{\partial P_t(c,\Y)}{\partial \Y},
\eeq
which can be solved through Laplace transform. Denoting $\hat P_t(c,\gamma)= \int d\Y P_t(c,\Y) e^{-\gamma \Y}$, we obtain
\beq
\frac{\partial \hat P_t(c,\gamma)}{\partial t} = \sum_{c'}  \hat P_t(c',\gamma) L_t^{ h_t}(c',c) -\dot h_t  \gamma \frac{\partial \psi_t(c,h_t)}{\partial h_t} \hat P_t(c,\gamma).
\eeq
Thus, the equation satisfied by $P_t(c,\gamma = 1)$ is identical with the equation Eq.~(\ref{pi_evo2}) satisfied by $\pi_t(c,h_t)$. Furthermore, since $P_0(c,\Y)=P_0(c) \delta(\Y)$, the two functions have identical initial conditions $\hat P_0(c,1)= \int d\Y \; P_0(c,\Y) \exp(-\Y)=P_0(c)=\pi_0(c,0)$. Therefore, these two functions must be identical, in other words: $\hat P_t(c,1)=\pi_t(c,h_t)$. Using the definition of the Laplace transform, it follows from this equality that
\beq
\pi_t(c,h_t)= \langle \delta(c-c_t) e^{-\Y_t}) \rangle_{[h_t]} \label{variante_sasa},
\eeq
an equation which can be called a Feynman-Kac formula \cite{Hummer2001_vol98,Liu2010_vol43}.

By multiplying Eq.~$\ref{variante_sasa}$ by an arbitrary observable $A_t(c,h_t)$ and integrating over $c$, one
obtains the following generalization of the Hatano-Sasa relation
\beq
\langle A_t(c_t,h_t) e^{-\Y_t} \rangle_{[h_t]} = \int dc \; \pi_t(c,h_t) A_t(c,h_t)= \langle A_t(c_t,h_t) \rangle_{\pi_t}, \label{variante_sasa2}
\eeq
where in the last equality $\langle ..\rangle_{\pi_t}$ denotes the average with respect to $\pi_t(c,h_t)$.
In the particular case that the initial condition is stationary and provided that the perturbed dynamics at constant perturbation $h$ is time independent, the Hatano-Sasa relation \cite{Hatano2001_vol86} is recovered from Eq.~\ref{variante_sasa2} in the particular case of a constant observable $A_t=1$.

We now consider a small variation with respect to the perturbation $h_t$. From the definition of $\Y_t$ in Eq.~\ref{def Y}, it follows that this quantity is small, at least of order one in $h_t$, and thus $e^{-\Y_t} \simeq 1-\Y_t$. Therefore, we have
\bea
\langle A_t(h_t) \rangle_{\pi_t} &\simeq& \langle A_t(c_t,h_t) \rangle_{[h_t]} - \left\langle \int_0^t \dot h_\tau  \partial_{h_\tau} \psi_\tau(c_\tau,h_\tau) A_t(c_t,h_t)  d\tau   \right\rangle_{[h_t]}, \\
& \simeq & \langle A_t(c_t,h_t) \rangle_{[h_t]} - \int_0^t d\tau  \dot h_\tau \left\langle  \left. \partial_{h} \psi_\tau(c_\tau,h)\right|_{h \rightarrow 0} A_t(c_t,h_t) \right\rangle, \label{dev_sasa}
\eea
where in the last equation, we have approximated the derivative with respect to $h_\tau$ by a
derivative with respect to $h$, an approximation which is correct to first order with respect to the perturbation, and at the same order in perturbation we have replaced the perturbed average by an unperturbed one.
Taking into account that the functional derivative of the l.h.s. of Eq.~\ref{dev_sasa} with respect to $h_{t'}$ vanishes for $t'<t$, and rewriting the second term of the r.h.s using an integration by parts, we obtain Eq.~\ref{reponse}.

\subsection{Interpretation of MFDT in terms of trajectory entropy}
\label{sec:trajec entropy}
In section \ref{definitions}, we have introduced a key quantity namely $\psi_t(c,h)=-\ln \pi_t(c,h)$. When properly evaluated on a specific trajectory $[c_t,h_t]$, the function $\psi_t(c_t,h_t)$ gives access to the functional $\Y_t$ defined in Eq.~\ref{def Y}, and when evaluated at small constant $h$ on a trajectory $[c_t]$, it allows to calculate the response in the MFDT according to Eq.~\ref{reponse}.
Clearly, this quantity must be closely related to the stochastic entropy introduced in Ref.~\cite{Seifert2005_vol95}.
Indeed, the stochastic system entropy is defined as
\beq
s_t(c_t,[h_t])=-\ln P_t(c_t,[h_t]),
\label{gen stoch entropy}
\eeq
and therefore depends functionally on the perturbation $[h_t]$. In contrast to that, the system entropy which is needed here is evaluated using a constant perturbation $[h]$ ,
\beq
s_t(c_t,[h])=-\ln P_t(c_t,[h])=-\ln \pi_t(c_t,h)=\psi_t(c_t,h).
\eeq

We now focus on the trajectories taken by the system, which can
be described by a set of discrete values $\mathcal{C}=\{c_0,c_1..c_N \}$, with the convention that the system is in state $c_0$ at time 0 and in state $c_N$ at time $t$. Furthermore, the transition from state $c_{j-1}$ to state $c_j$ occurs at the jumping times $\tau_j$. The stochastic system entropy can be decomposed as $s_t(c_t,[h_t])= - s_t^r(c_t,[h_t])+ s_t^{tot}(c_t,[h_t])$, in terms of the reservoir entropy production $s_t^r(c_t,[h_t])$ (also called medium entropy in Ref.~\cite{Seifert2005_vol95}) and the total entropy production $s_t^{tot}(c_t,[h_t])$.

The system entropy is a state function, which means that
\beq
\Delta s_t(c_t,[h_t])= -\ln P_t(c_t,[h_t])+\ln P_0(c_0,h_0).
\label{s state function}
\eeq
In contrast to that, the reservoir entropy and the total entropy are not state functions, but are trajectory dependent quantities, which can be written
\bea
\!\!\!\!\!\! \Delta s_t^{r}(c_t,[h_t]) &=& \sum_{j=1}^N \ln \frac{ w^{h_j}_{\tau_j}(c_{j-1},c_j)}{w^{h_j}_{\tau_i}(c_j,c_{j-1})}, \label{entropies} \\
\!\!\!\!\!\! \Delta s^{tot}_t(c_t,[h_t])&=&\sum_{j=1}^N \ln \frac{P_{\tau_j}(c_{j-1},[h_t])w^{h_j}_{\tau_j}(c_{j-1},c_j)}
{P_{\tau_j}(c_j,[h_t])w^{h_j}_{\tau_j}(c_j,c_{j-1})} - \int_0^t d\tau ( \partial_\tau \ln P_\tau ) (c_\tau,[h_\tau]), \nonumber
\eea
where $h_j$ is the value of the control parameter at the jump time $\tau_j$.
Between the jumps, $s_t^r$ is a constant function of the time while $s_t$ and $s_t^{tot}$ are in general non-constant but continuous functions of the time. All these functions, $s_t$, $s_t^r$ and $s_t^{tot}$ are discontinuous at the jump times $\tau_j$.

When adapted to the case of a constant perturbation $[h]$, the above decomposition of the system entropy leads to two terms in the MFDT. Starting from Eq.~\ref{reponse} together with Eq.~\ref{s state function}, one obtains
\beq
R(t,t')  =   - \frac{d}{dt'} \left\langle \left. \partial_h \Delta s_{t'}(c_{t'},[h]) \right|_{h \rightarrow 0}  A_t(c_t) \right\rangle = R_{eq}(t,t') - R_{neq}(t,t'),  \label{new reponse1}
\eeq
where
\beq
R_{eq}(t,t')= \frac{d}{dt'} \left\langle \left. \partial_h \Delta s_{t'}^r(c_{t'},[h]) \right|_{h \rightarrow 0}  A_t(c_t) \right\rangle, \label{R1}
\eeq
and
\beq
R_{neq}(t,t')=  \frac{d}{dt'} \left\langle \left. \partial_h \Delta s_{t'}^{tot}(c_{t'},[h]) \right|_{h \rightarrow 0}  A_t(c_t) \right\rangle.  \label{R2}
\eeq
This decomposition of the MFDT contains two terms, the first term $R_{eq}(t,t')$ which is analogous to the equilibrium FDT in a sense made more precise below, and the second term $R_{neq}(t,t')$ which represents an additive correction. Such a decomposition has been discussed by several authors following the original work of Ref.~\cite{Speck2006_vol74}. Note that the interpretation of the MFDT or standard FDT in terms of trajectory entropies is more recent \cite{Baiesi2009_vol103,Baiesi2009_vol137}; in the context of non-equilibrium stationary states this has been done in \cite{Verley2011_vol93,Seifert2010_vol89}. The present decomposition is very similar to that case, but here the total entropy production replaces the so-called adiabatic entropy production \cite{Esposito2007_vol76}, because the non-adiabatic part is non-zero and contributes to the second term $R_{neq}(t,t')$.

\subsubsection*{Reformulation of the MFDT using local currents}
Let us rewrite more explicitly the two terms above of the MFDT without the time derivatives present in Eqs.~\ref{new reponse1}-\ref{R2} but instead using local currents.
For the first term, using the expression for the reservoir entropy of Eq.~\ref{entropies}, we have
\begin{equation}
\hspace{-2cm}  \left\langle \Delta s_{t'}^{r}(c_{t'},\left[h\right])A_t(c_{t})\right\rangle =\int_{0}^{t'}ds\sum_{c,c',c''} \rho_{s}(c) w_{s}(c,c') \ln\left[\frac{w_{s}^{h}(c,c')}{w_{s}^{h}(c',c)}\right] \times \rho(c''t|c's) A_t(c''),
 \end{equation}
which implies that
\begin{eqnarray}
\hspace{-2.3cm} \frac{d}{dt'}\left\langle \partial_{h}\left.\Delta s_{t'}^{r}(c_{t'},\left[h\right])\right|_{h \ra 0}A(c_{t})\right\rangle  & = & \sum_{c,c',c''} \rho_{t'}(c) w_{t'}(c,c')\partial_{h}\left.\ln \frac{w_{t'}^{h}(c,c')}{w_{t'}^{h}(c',c)} \right|_{h \ra 0} \rho(c''t | c't') A_t(c''), \nonumber \\
 & \equiv & \left\langle j_{t'}(c_{t'})A_t(c_{t})\right\rangle = R_{eq}(t,t'),
 \end{eqnarray}
where $j_{t'}$ denotes the local current given by
\begin{equation}
j_{t'}(c')=\sum_{c}\frac{\rho_{t'}(c)}{\rho_{t'}(c')}w_{t'}(c,c')\partial_{h}\left.\ln \frac{w_{t'}^{h}(c,c')}{w_{t'}^{h}(c',c)} \right|_{h \ra 0}.
\label{courant j}
 \end{equation}
A property of this local current is that its average represents a physical current:
\beq
\left \langle j_{t'}(c_{t'}) \right \rangle = \sum_{c,c'} J_{t'}(c',c) \partial_h \ln \left. w_{t'}^{h}(c',c)\right|_{h \ra 0}, \label{average J}
\eeq
with $J_{t'}(c',c)=\rho_{t'}(c')w_{t'}(c',c) -\rho_{t'}(c)w_{t'}(c,c')$, the unperturbed probability current between the states $c$ and $c'$.

The same strategy can be used to transform the second term in the MFDT.
\bea
\hspace{-2.3cm} \left\langle  \Delta  s_{t'}^{tot}(c_{t'},\left[h\right])A_t(c_{t})\right\rangle & = & \int_{0}^{t'}ds\sum_{c,c',c''} \rho_{s}(c) w_{s}(c,c') \ln\left[\frac{\pi_s(c,h) w_{s}^{h}(c,c')}{\pi_s(c',h) w_{s}^{h}(c',c)}\right] \rho(c''t|c's) A_t(c'') \nonumber \\
& + & \int_0^{t'} d\tau \sum_{c',c''} \rho_\tau(c') (\partial_\tau \psi_\tau) (c',h) \rho(c''t|c' \tau) A_t(c''),
 \eea
which implies
\begin{eqnarray}
\hspace{-2.7cm}\frac{d}{dt'}\left\langle \partial_{h}\left.\Delta s_{t'}^{tot}(c_{t'},\left[h\right])\right|_{h \ra 0}A(c_{t})\right\rangle  & = & \sum_{c} \langle \frac{\rho_{t'}(c)}{\rho_{t'}(c_{t'})} w_{t'}(c,c_{t'})\partial_{h}\left.\ln \frac{\pi_{t'}(c,h) w_{t'}^{h}(c,c_{t'})}{\pi_{t'}(c_{t'},h) w_{t'}^{h}(c_{t'},c)} \right|_{h \ra 0} A_t(c_t) \rangle  \nonumber \\
 & + &  \left\langle \partial_h \left. (\partial_{t'} \psi_{t'})(c_{t'},h) \right|_{h \ra 0} A_t(c_{t}) \right\rangle.
 \end{eqnarray}
One can rewrite this in a more compact form
in terms of another local current $\nu_{t'}$ such that
\begin{equation}
\frac{d}{dt'}\left\langle \partial_{h}\left.\Delta s_{t'}^{tot}(c_{t'},\left[h\right])\right|_{h \ra 0}A(c_{t})\right\rangle \equiv\left\langle \nu_{t'}(c_{t'})A_t(c_{t})\right\rangle = R_{neq}(t,t'),
\label{Rneq}
 \end{equation}
where
\begin{equation}
\nu_{t'}(c')=\sum_{c}\frac{J_{t'}(c',c)}{\rho_{t'}(c')}\partial_{h}\left. \ln w_{t'}^{h}(c',c) \right|_{h \ra 0}.
\label{nu}
\end{equation}
Note that this equation together with Eq.~\ref{average J}, imply that both currents $\nu_{t'}(c')$ and $j_{t'}(c_{t'})$ have the same average.
In the end, the MFDT takes the form :
\beq
R(t,t')=\left\langle (j_{t'}(c_{t'})-\nu_{t'}(c_{t'})) A_t(c_t)\right\rangle.
\label{MFDT 2 terms}
\eeq
We emphasize that the decomposition of Eqs.~\ref{courant j}-\ref{nu} is a general result, which does not rely on any assumption about the form of the rates. Note that the response function can thus be written in a way which does not contain a time derivative provided two local currents $j_{t'}$ and $\nu_{t'}$ are introduced. This comes at the price that there is not a unique decomposition of this type \cite{Seifert2010_vol89}.

When additional assumptions are available about the transition rates, the two terms can be further transformed. One such assumption is a generalized detailed balance relation of the form :
\beq
\frac{w^{h_t}_t(c,c')}{w^{h_t}_t(c',c)}=\frac{w_t(c,c')}{w_t(c',c)}\exp (h_t d_t(c,c')), \label{bilan1}
\eeq
where $d_t(c,c')$ represents the variation of a physical quantity conjugate to $h_t$ during the transition from state $c$ to $c'$ at time $t$. In this case, the local current $j_{t'}$ can be simplified as :
\begin{equation}
j_{t'}(c')=\sum_{c}\frac{\rho_{t'}(c)}{\rho_{t'}(c')}w_{t'}(c,c')d_t(c,c').
\label{simplified j}
 \end{equation}
One particular usual choice of transition rates compatible with Eq.~\ref{bilan1} is
\beq
w^{h_{t'}}_{t'}(c,c')=w_{t'}(c,c') \exp{\left( \beta h_{t'} \frac{O(c')-O(c)}{2} \right)}, \label{bilan2}
\eeq
where $O(c)$ represents a physical time independent observable, and $d(c,c') = O(c') - O(c)$ its variation between state $c$ and $c'$. See \cite{Chetrite2011_vol143} for a general discussion on the various forms of perturbations in connection with the potential theory.

The reservoir entropy introduced in Eq.~\ref{entropies} only needs to be evaluated for a constant perturbation $[h]$, therefore
\bea
\hspace{-0.5cm}\partial_h \left. \Delta s_t^{r}(c_t,[h]) \right|_{h \ra 0} & = & \sum_{j=1}^N \partial_h \left. \ln \frac{ w^{h}_{\tau_i}(c_{j-1},c_j)}{w^{h}_{\tau_i}(c_j,c_{j-1})} \right|_{h \ra 0}= \beta  \sum_{j=1}^N \left( O(c_j)-O(c_{j-1}) \right), \nonumber \\
& = & \beta  \left( O(c_N)-O(c_0) \right) = \beta  \left( O(c_t) -O(c_0) \right).
\eea
Substituting this in the first term of the MFDT, namely, Eq.~\ref{R1}, one obtains
\beq
R_{eq}(t,t')= \beta \frac{d}{dt'} \left\langle O(c_{t'}) A(c_t)  \right\rangle,
\label{new R1}
\eeq
which is a form similar to the equilibrium FDT. Indeed, in the case of a perturbation around an equilibrium state, the average $\langle .. \rangle$ becomes an equilibrium average, and the equilibrium form of FDT is recovered.
We explain in the appendix A how the second term $R_{neq}(t,t')$ can be transformed in the continuous space limit for the particular case of a nearest-neighbour random walk.

In the end, the present framework provides a way to interpret the MFDT at the level of trajectory entropies in the case that a system is perturbed near an arbitrary non-equilibrium state. Thus, this framework generalizes the results obtained in Ref.~\cite{Verley2011_vol93} for non-equilibrium stationary states.

\section{Illustrative examples}
While the derivations above concern discontinuous pure jump Markov processes, the results are more general and their proofs are transposable for continuous diffusion processes \cite{Risken1989_vol}. For this reason, we provide in the following two illustrative examples of each kind: the first one corresponds to a continuous process of the Langevin type while the second one corresponds to a discrete jump process (Glauber dynamics).

\subsection{A particle obeying Langevin dynamics in an harmonic potential with time dependent stiffness and submitted to a quench of temperature}
\label{ex: Langevin}
We consider a particle in one dimension and in an harmonic potential obeying Langevin dynamics:
\begin{equation}
\dot x_t = - \frac{k_t}{\gamma} x_t + \frac{h_t}{\gamma} + \eta_t \; \mbox{ with } \; \left \langle \eta_t \eta_{t'}  \right \rangle = \frac{2T_t}{\gamma} \delta (t-t') \; \mbox{ and } \; \left \langle \eta_t \right \rangle = 0,
\label{Langevin}
\end{equation}
where $\eta_t$ is a Gaussian white noise, $k_t$ a time dependent spring constant, $\gamma$ a friction coefficient and $T_t$ the time dependent temperature of the bath, which starts from $T_0$ at $t=0$ and ends at $T_{t_f}$ at $t=t_f$. As a result of the non-stationary bath and of the time dependent spring constant, the system at time $t>0$ is not in equilibrium although it is assumed to be at equilibrium at $t=0$. We denote by $h_t$ an additional external perturbing force. For this system, one can compute explicitly, provided that the spring constant is integrable on interval $[0,t]$, the position at time t,
\begin{equation}
x_t = x_0 e^{-\int_0^{t}d \tau k_\tau / \gamma} +  e^{-\int_0^{t}d \tau k_\tau / \gamma} \int_0^t d\tau \left ( \frac{h_\tau}{\gamma} + \eta_\tau \right ) e^{\int_0^{\tau}d \tau' k_{\tau'} / \gamma},
\label{position}
\end{equation}
which is also a random Gaussian variable thanks to the linearity in $\eta$ and because the probability distribution of $x_0$ is the equilibrium Gaussian one at $T_0$, i.e $  \rho_0(x_0) = \exp(-k_0x_0^2/(2T_0))/Z) $ with $Z$ the partition function. From Eq.~ \ref{position}, we obtain
\bea
\mu_t &=& \langle x_t \rangle_{[h_t]} = \int_0^t d\tau  \frac{h_\tau}{\gamma} \exp{\left ( - \int_{\tau}^{t}d \tau' k_{\tau'} / \gamma \right ) }, \label{moyenne_xt}  \\
\s_t^2 &=&\langle x_t^2 \rangle_{[h_t]} - \langle x_t\rangle_{[h_t]}^2, \\
 &=& \langle x_0^{2}\rangle \exp \left( -2 \int_0^{t}d \tau k_\tau / \gamma \right) +\int_0^t d\tau  \frac{2T_\tau}{\gamma} \exp \left ( - 2 \int_{\tau}^{t}du k_{u} / \gamma \right ).
\eea
The functional derivative of Eq.~\ref{moyenne_xt} with respect to $h_{t'}$ gives directly the response function
\begin{equation}
R(t,t') = \left. \frac{\delta \langle x_t \rangle_{[h_t]} }{\delta h_{t'}} \right|_{h \rightarrow 0}= \frac{1}{\gamma} \exp \left ( -  \int_{t'}^{t}d\tau k_{\tau} / \gamma \right ). \label{Rep_langevin}
\end{equation}
As we show below, this result can also be recovered from the MFDT of Eq.~\ref{reponse}. Since $x_t$ is Gaussian variable, we deduce from this that the probability density function at time $t$ is
\beq
\hspace{-1cm}P_t(x,[h_t]) = \frac{1}{(2 \pi \s_t^2)^{1/2}}\exp \left [ \frac{-1}{2\s_t^2} \left (x-\int_0^t d\tau \frac{h_\tau}{\gamma}   \exp{\left ( - \int_{\tau}^{t}d \tau' k_{\tau'} / \gamma \right ) }\right )^2 \right  ].
\eeq
We note that $\s_t^2$ does not depend on the perturbation $[h]$ but $\mu_t$ does. To obtain the probability density function $\pi_t(x,h)$, we just make the perturbation constant using  $\pi_t(x,h) = P_t(x,[h]) $, so:
\begin{eqnarray}
\pi_t(x,h) &=& \frac{1}{(2 \pi \s_t^2)^{1/2}}\exp \left [ \frac{-1}{2\s_t^2} \left (x-\frac{h}{\gamma}\int_0^t d\tau   \exp{\left ( - \int_{\tau}^{t}d \tau' k_{\tau'} / \gamma \right ) }\right )^2 \right  ].
\end{eqnarray}
Note that $\pi_t(x,h_t)$ is indeed distinct from $P_t(x,[h_t])$ as emphasized from the beginning. Since we now have the key function $\pi_t$, we can compute the response function and the functional $\Y_t$.

To obtain the response, we calculate $\psi_t(x,h) = - \ln \pi_t(x,h)$ and its derivative with respect to $h$:
\beq
\partial_{h} \psi_t(x,h) = -\frac{1}{\s_t^2}(x-h I_t)I_t,
\label{deriv_psi}
\eeq
where
\beq
I_t = \frac{1}{\gamma} \int_0^t d\tau  e^{\left ( - \int_{\tau}^{t}d \tau' k_{\tau'} / \gamma \right ) },
\eeq
resulting in
\begin{equation}
\left.  \langle \partial_{h} \psi_{t'}(x_{t'}) x_t  \rangle \right |_{ h\ra 0} = - \frac{I_{t'}}{\s_{t'}^2}  \left. \langle x_t x_{t'} \rangle \right |_{ h\ra 0}.
\end{equation}
This last expression requires the two times correlation function for $x$ that can be obtained from Eq.~\ref{position}, assuming $t'<t$
\beq
\langle  x_{t'} x_t \rangle = \mu_t \mu_{t'} +  \s_{t'}^2 \exp \left ( - \int_{t'}^t d\tau k_\tau /\gamma \right   ).
\label{corr_x}
\eeq
Since that $ \mu_t = \mu_{t'} = 0 $ in the limit $h \longrightarrow 0$, we obtain
\bea
\left. \langle \partial_{h} \psi_{t'}(x_{t'})  x_t  \rangle \right |_{ h\ra 0} &=& - I_{t'}\exp \left ( - \int_{t'}^t d\tau k_\tau /\gamma \right), \label{hint1} \\
&=& - \frac{1}{\gamma} \int_0^{t'} d\tau  e^{\left (\int_{0}^{\tau}d \tau' k_{\tau'} / \gamma \right ) } \exp \left ( - \int_{0}^t d\tau k_\tau /\gamma \right). \label{hint2}
\eea
Now, by taking the opposite of the time derivative with respect to $t'$, we obtain the response function of Eq.~\ref{Rep_langevin}, which provides a verification of Eq.~\ref{reponse} on this particular example. Note that the response function does not have time translational symmetry (it is not solely a function of the time difference $t-t'$) in the general case that the spring constant is time-dependent. Furthermore, this response function has the property of being independent of the protocol of temperature variation.

\graph{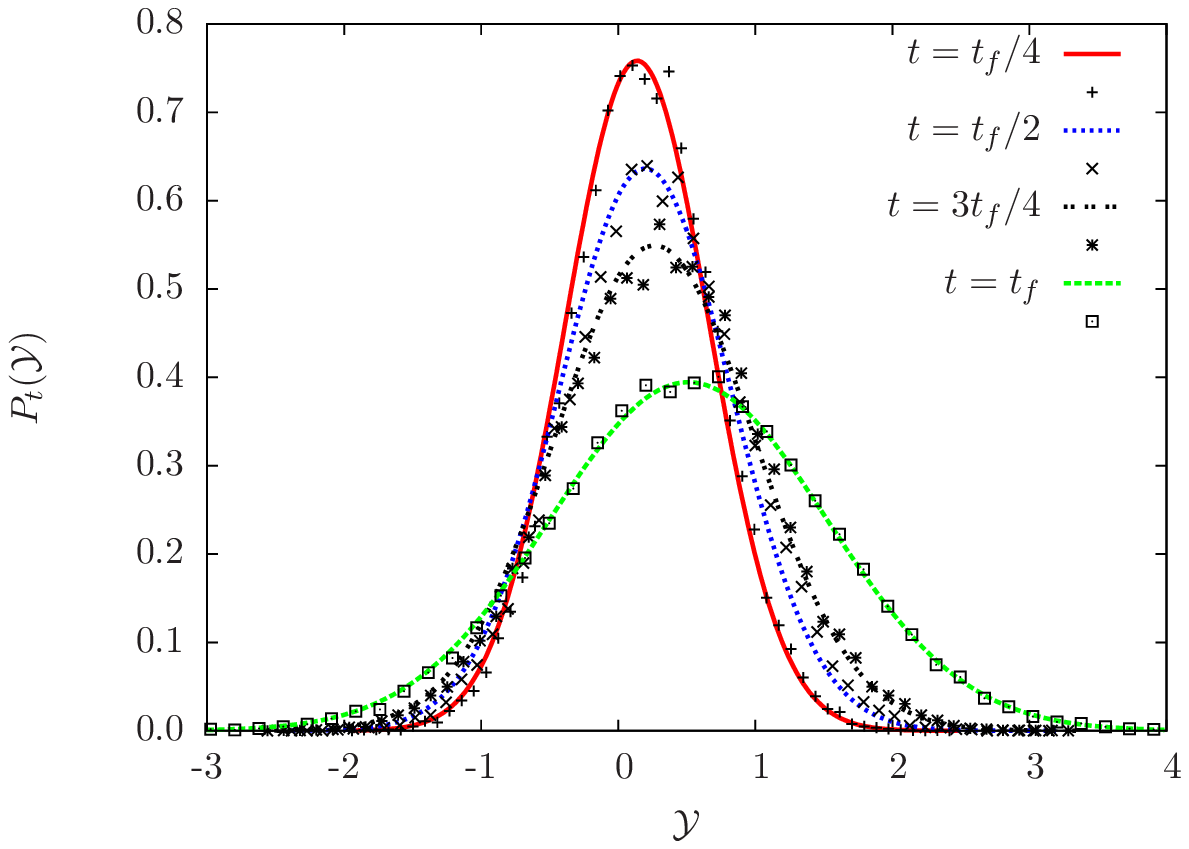}{Probability density functions $P_t(\Y)$ at five different times $t$ for a  particle in an harmonic trap obeying Langevin dynamics. The symbols represent an estimate of $P_t(\Y)$ based on $10^5$ trajectories of total duration $t_f=5.12$; the solid line is the Gaussian probability density which has the mean and variance given respectively by Eq.~\ref{moyY} and \ref{varY}.
The system is at $t>0$ in a non-equilibrium state due to an imposed time-dependent spring constant $k_t=5 + 2.5\sin(\frac{\pi t}{t_f})$, and a time-dependent heat bath temperature, which is such that it is $T_0 = 5$ at $t=0$ and $T_t=1$ for $t>0$. The friction coefficient is $\gamma = 1$. This system is further perturbed by a force, according to the protocol $h_t=5\sin(\frac{\pi t}{t_f})$.}{fig1}

\graph{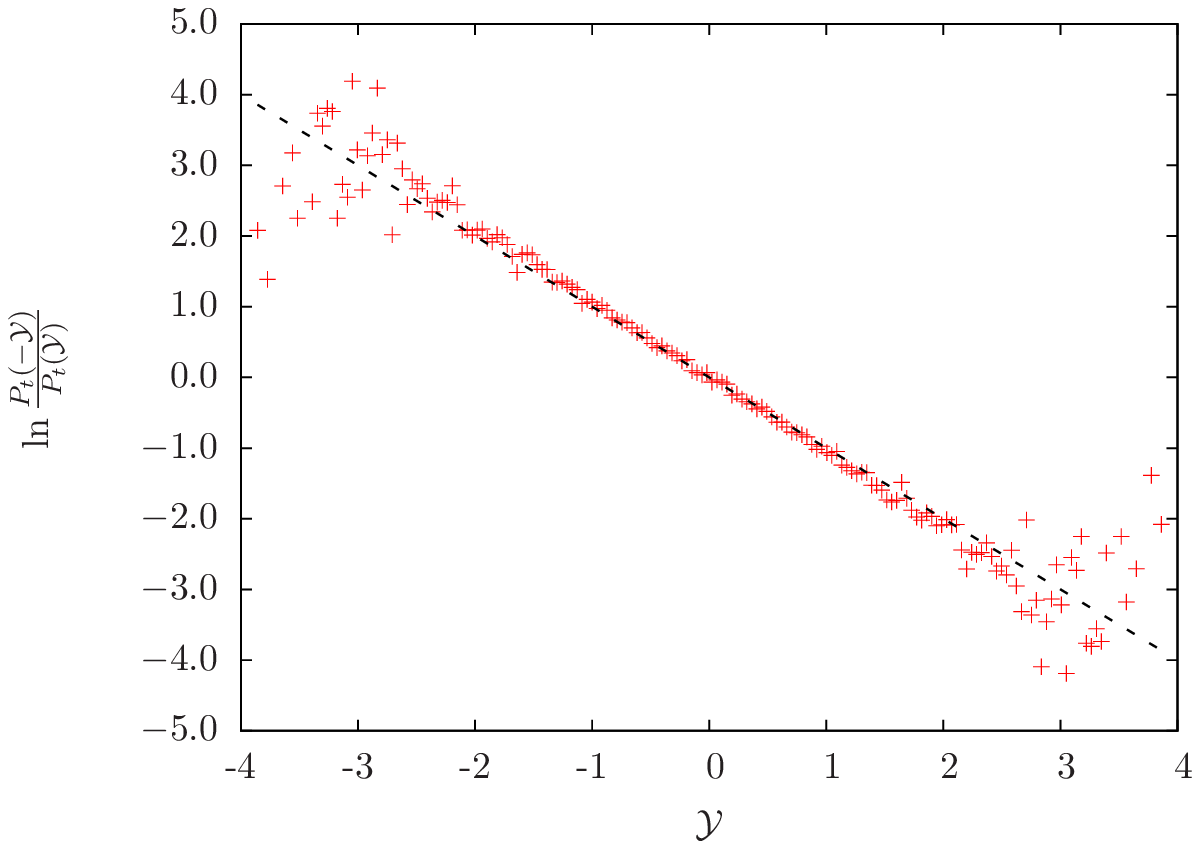}{Numerical test of the detailed fluctuation relation of Eq.~\ref{DFR simple} satisfied by the probability distribution of $\Y$. Here $t=t_f=5.12$, and all the other parameters are the same as in Figure \ref{fig1}.}{fig2}

We obtain $\Y_t$ from Eq.~\ref{def Y} and Eq.~\ref{deriv_psi},
\beq
\Y_t = - \int_0^t d\tau \frac{\dot{h}_\tau I_\tau}{\s_\tau^2}(x_\tau-h_\tau I_\tau),
\label{estimate of Y}
\eeq
which is linear in $x_\tau$ and hence is also a Gaussian variable. Its mean value and variance of $\Y_t$ are respectively
\bea
\langle \Y_t \rangle &=& \int_0^t d\tau \frac{\dot{h}_\tau I_\tau}{\s_\tau^2}(h_\tau I_\tau - \mu_\tau), \label{moyY}  \\
\langle \Y_t^2 \rangle &-& \langle \Y_t \rangle^2 = \int_0^t d\tau \int_0^t d\tau' \frac{\dot{h}_\tau I_\tau\dot{h}_{\tau'} I_{\tau'}}{\s^2_\tau \s^2_{\tau'}} \left( \langle x_\tau x_{\tau'} \rangle - \mu_\tau \mu_{\tau'} \right).
\label{varY}
\eea
In the end, after transforming Eq.~\ref{varY} using an integration by parts, we find that
$ \langle \Y_t^2 \rangle - \langle \Y_t \rangle^2 = 2 \langle \Y_t \rangle $. Since $\Y_t$ is Gaussian process, this relation implies that the probability density function $P_t(\Y)$ satisfies the following detailed fluctuation relation
\beq
\frac{ P_t(\Y)}{P_t(-\Y )}= \exp{(\Y)},
\label{DFR simple}
\eeq
which we have also confirmed through a numerical determination of the distribution of $\Y_t$ as shown in figures \ref{fig1}-\ref{fig2}.

Remarkably, this detailed fluctuation relation holds although no dual process has been invoked. Thus, this relation differs in an essential way from the generalized Crooks theorem given in \cite{Chetrite2011_vol143} (see Eq.~133). Another difference between Eq.~\ref{DFR simple} and the generalized Crooks theorem of that reference is that in the generalized Crooks relation, the initial condition in the forward or backward processes are taken according to the distribution $\pi_0(x,h_0)$ and $\pi_t(x,h_t)$ respectively. In contrast to this, in Eq.~\ref{DFR simple} the initial condition in which the system is prepared results from the application of unperturbed dynamics. For this reason, one may say that Eq.~\ref{DFR simple} is closer to a relation of the Bochkov and Kuzovlev type rather than to a Crooks relation \cite{Horowitz2007_vol}.

Finally, we would like to emphasize that Eq.~\ref{DFR simple} is a very general result for linear Langevin dynamics. We have checked in \ref{multidimensional} that Eq.~\ref{DFR simple} can indeed be extended to a  general multidimensional linear Langevin dynamics.

\subsection{The 1D Ising model with Glauber dynamics}
\subsubsection{Introduction}
We now move to a more complex system with many interacting degrees of freedom, which will allow for phase transitions and ordering phenomena absent from the previous example. The system is the Ising-Glauber chain in 1D, which when submitted to a temperature quench, is a paradigm for coarsening dynamics \cite{Glauber1963_vol4}.
For this system, explicit exact expressions of the correlation and response functions have been obtained; and the ratio between these two quantities admits a non-trivial limit, which is a universal quantity in the case of a quench to the critical temperature \cite{Godreche2000_vol33}. Multi point correlation functions have also been calculated analytically in order to test theoretical ideas about the dynamic heterogeneities of glasses \cite{Mayer2004_vol37,Mayer2005_vol2005}.
In more general spin systems, the correlations or response can not be obtained analytically, but the response function has been shown to be related to correlations characteristic of the non-perturbed system \cite{Lippiello2005_vol71,Diezemann2005_vol72,Chatelain2003_vol36,Ricci-Tersenghi2003_vol68}, a conceptual progress but also a definite advantage for numerical simulations as compared to previous methods.
In the following, we illustrate the framework of modified fluctuation-dissipation theorem presented in the previous sections for the Glauber-Ising chain submitted to a quench of temperature. Using analytical calculations, we first show that we can recover the known exact response function using this formalism. We then present some numerical simulations to confirm the theoretical expectations.

\subsubsection{Definition of the rates}
This Ising-Glauber chain is made of $L$ Ising spins $\s_i=\pm 1$ with $i=1..L$ in one dimension, and is described by the following Hamiltonian
\beq
\mathcal{H}(\{ \s \})= -J \sum_{i=1}^{L} \s_i \s_{i+1} -  H_m \s_m,
\eeq
where $J$ is the coupling constant and $H_m$ a magnetic field which acts on the spin $m$. We assume periodic boundary conditions. The magnetic field $H_m$ will be the only control parameter. In principle, we could allow for many control parameters corresponding to magnetic fields present on any lattice site, but since we are mainly interested in the linear response regime, we restrict ourselves to the case where this magnetic field only acts on the spin $m$.
The Ising chain is assumed to be initially in equilibrium at an infinite temperature. At $t=0$,  it is submitted to an instantaneous quench which brings the temperature to $T$, and the system in a non-equilibrium state. The system further evolves after the quench from the time 0 to the time $t'$ where the small magnetic field $H_m$ is turned on. Therefore we assume that $H_m(t)=H_m \theta(t-t')$, with $t>0$ and $\theta$ the Heaviside function.

The probability to find the system in the state $\{ \s \}= \{ \s_1,...,\s_L \}$ at time $t$, $P_t(\cf)$, obeys the following master equation
\beq
\frac{\partial P_t(\cf)}{\partial t} = - \sum_{i} w^{H_m}( \{ \s \},\{ \s \}^i ) P_t(\cf) + \sum_{i} w^{H_m}( \{ \s \}^i,\{ \s \} ) P_t(\cf^i), \label{master_spin}
\eeq
where $w^{H_m}( \{ \s \},\{ \s \}^i )$ is the rate to jump from the configuration $\{ \s \}$ to the configuration $\{ \s \}^i = \{ \s_1,...,-\s_i,...,\s_L \}$.

Following Ref.~\cite{Godreche2000_vol33}, we choose the rates in the presence or absence of a field to be given respectively by
\bea
w^{H_m}( \{ \s \},\{ \s \}^i ) &=& \frac{\alpha}{2}(1- \s_i \tanh(\beta J (\s_{i-1} + \s_{i+1}) + \beta H_m \delta_{im})), \label{taux} \\
w( \{ \s \},\{ \s \}^i ) &=& \frac{\alpha}{2}\left(1- \s_i \frac{\gamma}{2} (\s_{i-1} + \s_{i+1}) \right),
\label{taux_np}
\eea
where $\alpha$ denotes the inverse characteristic time scale of the transitions (which we take below to be equal to 1), $\gamma = \tanh(2\beta J)$ and $\beta=1/T$, the inverse of the temperature after the quench. Note that these rates depend on time (for $t>0$) only via the control parameter $H_m$. It is important to also realize that this form of the rates is just one of the possible choices compatible with the detailed balance condition, which imposes that
\beq
\frac{ w^H(\{ \s \},\{ \s' \})}{w^H(\{ \s' \},\{ \s \})}=\frac{e^{- \beta \mathcal{H}(\{ \s' \} )}}{e^{- \beta \mathcal{H}(\{ \s \})}}.
\label{DB}
\eeq
Other forms are possible, for instance Eq.~\ref{bilan2} corresponds to a different acceptable choice, which is the one made in Ref.~\cite{Lippiello2005_vol71} while Eqs.~\ref{taux}-\ref{taux_np} is the choice of Refs.~\cite{Godreche2000_vol33,Ricci-Tersenghi2003_vol68}.

\subsubsection{Analytical verification of the MFDT}
Unlike in the previous example of a particle obeying Langevin dynamics, where the non-stationary probability distribution $\pi_t$ was analytically solvable even in the presence of the perturbation due to the assumption of an harmonic potential, in the present problem, the probability distribution $\pi_t(\cf,H_m)$ to find the system in an arbitrary configuration $\cf$ at time $t$ with a constant magnetic field $H_m$ applied from $t=0$ is difficult to obtain even in the absence of a magnetic field ($H_m=0$).
As we show below, this distribution is not required to evaluate the response function for a general one spin dependent observable $A(\s_n)$, because in this particular case only the reduced one-spin distribution $\pi_t(\s_n,H_m)= \sum_{\cf \neq \s_n} \pi_t(\cf,H_m)$ is needed and fortunately, this reduced distribution can be calculated analytically. We define the response function as
\beq
R_{n-m}(t,t') = T \left. \frac{\delta \langle A(\s_n(t)) \rangle_{[H_m]}}{\delta H_m(t')}\right |_{H_m \ra 0},  \label{rep_spin}
\eeq
which contains an extra factor $T$ with respect to the definition used in previous sections. The reason for this extra factor is purely a matter of convenience, but historically it was introduced in order to provide the response function with a well defined limit when $T \rightarrow 0$.
According to the MFDT of Eq.~\ref{reponse}, this response function is
\bea
R_{n-m}(t,t') &=& T \frac{d}{dt'}  \!\!  \sum_{\s_n,\cfp} \!\!\!  A(\s_n) \rho(\s_nt|\cfp t') \rho_{t'}(\cfp) \! \left.\frac{\partial_{H_m} \pi_{t'}(\cfp,H_m)}{\pi_{t'}(\cfp,H_m)} \right |_{H_m \ra 0}, \nonumber \\
&=&  T \frac{d}{dt'}  \!\!  \sum_{\s_n,\cfp} \!\!\!  A(\s_n) \rho(\s_nt|\cfp t')  \! \left. \partial_{H_m} \pi_{t'}(\cfp,H_m) \right |_{H_m \ra 0}. \nonumber
\eea
where in the last step, we used $\pi_{t'}(\cfp,0)= \rho_{t'}(\cfp)$.

To progress, we need an explicit expression for the propagator $\rho(\s_n t| \cfp t')$, which is the probability to find the system with spin $n$ in state $\s_n$ at time $t$ in the unperturbed dynamics given that the system was in the state $\cfp$ at time $t'$. In fact, this propagator is directly related to the average magnetization at time $t$  when the system starts in the state $\cfp$ at time $t'$, namely
$\langle \s_n(t) \rangle_{\cf(t')=\cfp},$
\beq
\rho(\s_n t| \cfp t')=\frac{1}{2} \left( \s_n \langle \s_n(t) \rangle_{\cf(t')=\cfp} + 1 \right).
\eeq
From Eq.~\ref{master_spin}, one can show that this average magnetization
in the absence of an applied field satisfies \cite{Glauber1963_vol4}
\beq
\langle \s_n(t) \rangle_{\cf(t')=\cfp} =  \sum_k  G_{n-k}(t-t') \sigma_k',
\label{magnetization}
\eeq
where $G_k(t)$ is the Green function of the problem. In the following, we consider the thermodynamic limit $L \rightarrow \infty$,
 in which case, the Green function can be written as $G_{k}(t) = e^{-t} I_k(\gamma t)$, in terms of $I_k$ the modified Bessel function. It follows from the above two equations that
\bea
\rho(\s_n t|\cfp t')  &=& \frac{\s_n}{2}\sum_k \s_k' G_{n-k}(t-t') +\frac{1}{2}.   \label{propagateur}
\eea

After substituting this expression in Eq.~\ref{rep_spin}, one obtains
\bea
R_{n-m}(t,t')&=& T \frac{d}{dt'}  \sum_{\s_n,\cfp} \!\!\! A(\s_n) \left.\partial_{H_m} \pi_{t'}(\cfp,H_m) \right|_{H_m \ra 0} \nonumber \\
&& \times \left [ \frac{\s_n}{2}\sum_k \s_k' G_{n-k}(t-t') +\frac{1}{2} \right ].
\label{Rep_calc_intermediaire}
\eea
The fact that the observable $A(\sigma_n)$ only depends on a single spin $\sigma_n$ leads to a  simplification of the expression above since it is possible to sum over all spins in $\cfp$ except for the $k$th spin of that set. The second term in the bracket of Eq.~\ref{Rep_calc_intermediaire} vanishes due to the normalization condition
$\sum_{\cfp} \pi_{t'}(\cfp,H_m) = 1$; and we obtain
\begin{equation}
R_{n-m}(t,t')= T \frac{d}{dt'}  \sum_{k} \sum_{\s_k',\s_n}\frac{1}{2}  \s_k'\s_nA(\s_n)  G_{n-k}(t-t') \left.\partial_{H_m} \pi_{t'}(\s_k',H_m) \right |_{H_m \ra 0}.
\label{resp2}
\end{equation}
This equation shows that the response now only depends on the reduced one-spin distribution $\pi_t(\s_n,H_m)= \sum_{\cf \neq \s_n} \pi_t(\cf,H_m)$, which can be obtained analytically at first order in the applied field $H_m$ from the magnetization. Indeed, in the presence of a field, the magnetization at a time $t>0$ is
\beq
\langle \s_n(t) \rangle_{\cf(0)=\cfp}
\simeq \sum_k G_{n-k}(t) \sigma_k' + \beta H_m \int_0^{t} G_{n-m}(t-t') K(t') dt',
\label{magnetization2}
\eeq
where $\displaystyle K(t) = 1- \frac{\gamma^2}{2}(1+\langle \s_{n+1}(t)\s_{n-1}(t) \rangle)$, a correlation function which is known analytically.
For instance, in the case of a $T=0$ quench where $\gamma(t)=1$ for $t>0$, we have $K(t)= e^{-2t} (I_0(2t)+I_2(2t)+2I_1(2t))/2$ \cite{Krapivsky2010_vola}.

Since the chain is initially at infinite temperature, the average magnetization of spin $k$ at time $t=0$ vanishes for all $k$. It follows that the first term of Eq.~\ref{magnetization2} vanishes when the average over the initial condition is performed. Using Eq. \ref{propagateur}, we obtain  the distribution $\pi_t(\s_n,H_m)$ at first order in the applied field
\beq
\pi_t(\s_n,H_m) \simeq \frac{1}{2} + \frac{1}{2} \beta H_m\s_n \int_0^{t} G_{n-m}(t-t') K(t') dt'. \label{pi_spin}
\eeq
We can use this distribution to write the response function in Eq.~\ref{resp2} in a more explicit form. After summing over $\s_k'$, we have
\begin{eqnarray}
R_{n-m}(t,t') &=& \frac{d}{dt'}  \sum_{k,\s_n} \frac{1}{2} \s_n A(\s_n) G_{n-k}(t-t') \int_0^{t'} G_{k-m}(t'-u) K(u) du \nonumber \\
&=& \frac{A(1)-A(-1)}{2} \frac{d}{dt'} \int_0^{t'}  \sum_{k} G_{n-k}(t-t') G_{k-m}(t'-u) K(u) du \nonumber \\
&=& \frac{A(1)-A(-1)}{2} G_{n-m}(t-t') K(t'),
\label{final MFDT}
\end{eqnarray}
which agrees indeed with the response function obtained in \cite{Godreche2000_vol33} in the case $ A(\s_n)=\s_n $.

Besides recovering the known response function of the Glauber-Ising chain, we can also investigate the separate contributions of the two local currents $j_{t'}$ and $\nu_{t'}$ introduced in the section \ref{sec:trajec entropy}.
It is straightforward to show using the detailed balance condition of Eq.~\ref{DB} that the first term in the MFDT, namely $R_{eq}(t,t')$, can be written in a form similar to that of the equilibrium FDT (note that in the equation below, the index $j$ represents the discrete times when jumps occur):
\bea
\partial_{H_m} \left. \Delta s_t^{r}(\{ \s \}_t,[h]) \right|_{H_m=0} & = &   \sum_{j=1}^N \partial_{H_m} \left. \ln \frac{ w^{H_m}_{\tau_j}(\{ \s \}({j-1}),\{ \s \}({j}))}{w^{H_m}_{\tau_j}(\{ \s \}({j}),\{ \s \}({j-1}))} \right|_{H_m=0} \\ & = & \beta \sum_{j=1}^N \left( \s_m(j)-\s_m(j-1) \right), \\
& = & \beta \left(  \s_m(N)-\s_m(0) \right) = \beta \left(  \s_m(t)-\s_m(0) \right).
\label{equil part discrete time}
\eea
Substituting this in Eq.~\ref{R1}, one obtains the equilibrium part of the response for a general multi-spins observable $A( \{ \s \}_t)$,
\beq
R_{eq}(t,t')= \beta \frac{d}{dt'} \left\langle A( \{ \s \}_t) \s_m(t') \right\rangle.
\label{new R1 Ising}
\eeq
Alternately, one can also recover this result through an evaluation of the local current $j_{t'}$ using Eq.~\ref{courant j}.

For the term associated with the local current $\nu(t')$, one can show either from the decomposition of the stochastic entropy into two terms or from Eqs.~\ref{Rneq}-\ref{nu} that the expected part of the MFDT is recovered,
in other words that $R_{neq}(t,t')=R(t,t')-R_{eq}(t,t')$, where $R(t,t')$ is the response given in Eq.~\ref{final MFDT} for the case of a one-spin observable $A(\s_n)$. In the literature \cite{Lippiello2005_vol71,Diezemann2005_vol72}, the term $R_{neq}(t,t')$ is called the asymmetry, its precise form depends on the specific form of the rates (unlike the first term which only depends on the ratio of forward to backward rates) and it vanishes under equilibrium conditions. The present derivation makes also clear that there is in principle a physical observable associated with this term, namely $\Delta s_t^{tot}(\{ \s \}_t,[h])$.

\subsection{Numerical verification}

\graph{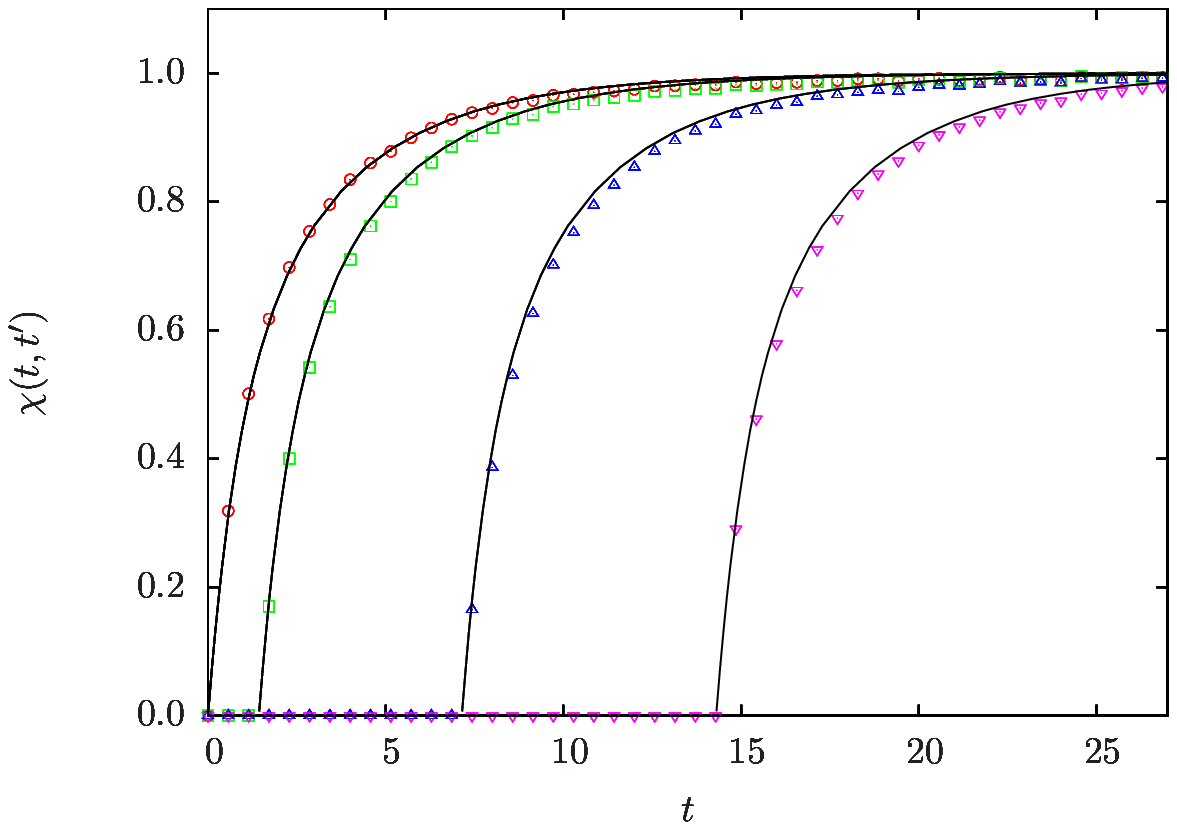}{Integrated response functions $\chi(t,t')$ versus time $t$ for a step protocol of magnetic field $H_m=0.05 $ starting at various values of the waiting time $t'$ after the initial quench at $t=0$. The different values of $t'$ are $t'=0$ for circles, $t'=1.43$ for squares, $t' =7.14$ for triangles and $t'=14.3$ for inverted triangles. The parameters are the following: $J=0.5$, $\alpha=1$, $T=1$ and $L =14$. The response is calculated on the same spin where the magnetic field is applied, here $n=m=3$. The averages have been done with $10^6$ trajectories of $400$ time steps of length $dt=0.07$. The continuous lines stand for the integrated response obtained analytically, while the symbols have been obtained from the MFDT of Eq.~\ref{reponse} using the unperturbed dynamics.}{fig3}

\graph{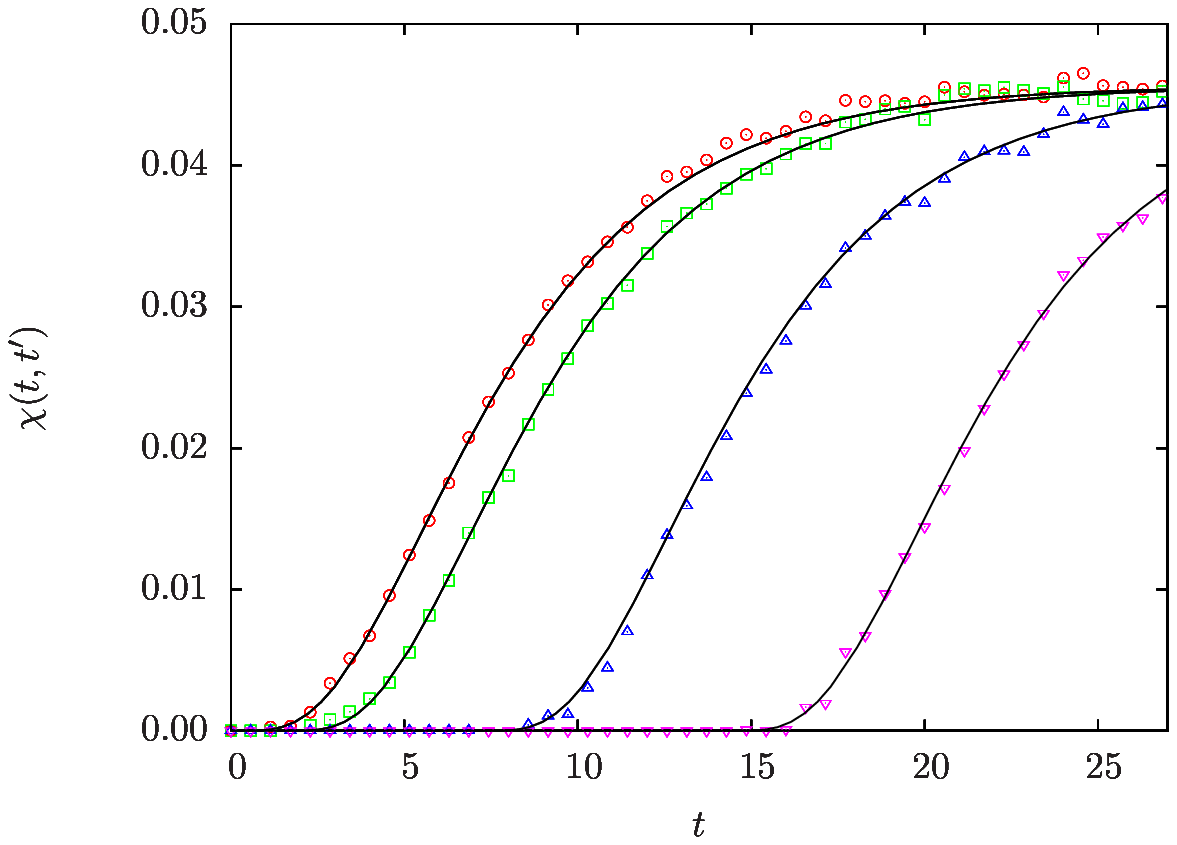}{Integrated response functions $\chi(t,t')$ versus time $t$ in the same conditions as in figure \ref{fig3} except that now the response is evaluated on a different spin ($n=7$) as compared to where the magnetic field is applied ($n=3$) and the averages have been done with $10^7$ trajectories.}{fig5}

As mentioned above, the distribution $\pi_t(\cf,H_m)$ does not seem to be accessible analytically. In order to test our framework, we have thus determined these distributions numerically from simulations for various values of $H_m$ and $\cf$. Then, we have calculated the response function $R(t,t')$ via Eq.~\ref{reponse} and using trajectories which were simulated according to the dynamics in the absence of a magnetic field. In figures \ref{fig3}-\ref{fig5}, the integrated response defined by
\beq
\chi_{n-m}(t,t')= \int_{t'}^t d\tau R_{n-m}(t,\tau),
\eeq
is shown, where the symbols represent the response function obtained from such simulations at zero field via Eq.~\ref{reponse} and
the solid line is the analytical expression obtained from Eq.~\ref{final MFDT}. Since Eq.~\ref{final MFDT} takes a simple form in Laplace space \cite{Godreche2000_vol33}, this solid line was obtained through a numerical inverse Laplace transform of that equation. 


In the simulations at zero field, we have used a small system size of $L=14$. It is difficult to go to significantly larger sizes with the present algorithm, because the numerical determination of the distribution $\pi_t(\cf,H_m)$ becomes rapidly a difficult task in large systems given that the configuration space grows as $2^N$.

\section{Conclusion}

In this paper, we have presented a framework which can be used to generalize the fluctuation-dissipation theorem to non-equilibrium systems obeying markovian dynamics. We have shown, using alternatively the linear response theory or a first order development of a fluctuation relation, that the main result of Ref.~\cite{Prost2009_vol103} for systems in a non-equilibrium stationary state is generalizable to systems which are near non-stationary states. This generalization is important because it restores a fluctuation dissipation theorem in the form of a unique correlation between physical observables and it opens many new possibilities to apply this framework to experiments.

In fact, this framework is applicable to systems which are prepared in a non-equilibrium state, and which are then be further probed through the application of a \emph{time-dependent} control parameter. This situation is typically the one encountered in studies of slow relaxing or aging systems, but it is also a frequent situation in biological systems. One outcome of our approach is that it is possible to replace this complicated problem by a somewhat simpler problem, namely the problem of determining the probability distribution $\pi_t(c,h)$ to find the system in a  non-equilibrium state $c$ but with a \emph{time-independent} perturbation $h$. Having to consider only a time-independent perturbation to probe a non-equilibrium system, should be a definite advantage both from an experimental and theoretical point of view.

Our study of a Brownian particle in an harmonic potential and submitted to a quench of temperature, raises the question of the validity of fluctuation relations for particles in contact with a nonequilibrium bath. This kind of studies may be important to understand the nonequilibrium fluctuations of a Brownian particle confined in a gel as in the experiment of Ref.~\cite{Gomez-Solano2011_vol106}.
We hope that our work will trigger further studies on the applications of stochastic thermodynamics to the characterization of non-equilibrium systems, and in particular for non-equilibrium systems which result from contact with a nonequilibrium bath or from a coarsening process.

\section*{Acknowledgement}
We acknowledge stimulating discussions with F. Krzakala and B. Wynants.

\appendix

\section{Link with previous formulations}

\subsection*{Equivalence between different expressions of the work like functional}

As mentioned in the main text, the functional $\Y_t$ defined in Eq.~\ref{def Y} has appeared before in Ref.~\cite{Chetrite2009_vol80}. We explain here how to make contact with the different notations. The functional denoted here $\Y_t$, corresponds to the one called $W_t$  in Eq.~24 of this reference. In order to see this correspondence, one should choose the quantity defined as $f_t$ in Ref.~\cite{Chetrite2009_vol80} to be equal to $\pi_t(c,h_t)$. With this choice,
\bea
 \hspace{-2cm} W_{t} & \equiv & \int_{0}^{t}d \tau \left(\pi_{\tau}(\cdot,h_{\tau})^{-1}\left(L_{\tau}^{h_{\tau}}
\right)^{\dagger}\left[\pi_{\tau}(\cdot,h_{\tau})\right]-\frac{\partial}{\partial \tau}\left(\ln\pi_{\tau}(\cdot,h_{\tau})\right)\right)(c_{\tau}) \\
\hspace{-2cm} & = & \int_{0}^{t}d\tau\left(\pi_{\tau}(c_{\tau},h_{\tau})^{-1}\left(\partial_{\tau}
 \pi_{\tau}\right)(c_{\tau},h_{\tau})-\left(\partial_{\tau}\ln\pi_{\tau}\right)
 (c_{\tau},h_{\tau})-\dot{h}_{\tau}\partial_{h_{\tau}}\ln\pi_{\tau}
 (c_{\tau},h_{\tau})\right) \nonumber \\
 \hspace{-2cm} & = & -\int_{0}^{t}d\tau\dot{h}_{\tau}\partial_{h_{\tau}}\ln\pi_{\tau}(c_{\tau},h_{\tau}),
 \eea
 where in the first line, the dagger stands for adjoint operator and the $\cdot$ indicates that the expression is to be understood as matrix product before evaluation on trajectory $c_\tau$. This calculation shows that $W_t$ indeed coincides with $\Y_t$ defined in Eq.~\ref{def Y}.


\subsection*{Equivalence between the discrete and continuous formulation of the MFDT}

In Refs.~\cite{Chetrite2008_vol,Chetrite2009_vol80}, a modified fluctuation-dissipation has been derived for general continuous diffusion processes. Here we show that the present framework formulated for discrete jump processes leads to the same results, when the appropriate continuous limit of the master equation is taken \cite{VandenBroeck2010_vol82}.

We shall assume the same parametrization of the rates as that given in Eq.~\ref{bilan2}. The equilibrium contribution in the response, $R_{eq}(t,t')$, can be written in the same way as in Ref.~\cite{Chetrite2009_vol80}, therefore, we shall focus on the other non-equilibrium contribution namely $R_{neq}(t,t')$. For simplicity, let us consider a nearest-neighbour random walk on a 1D lattice, in which a random walker at position $m$ can only jump to neighbouring sites $m \pm 1$. We denote the actual distance between all the sites by $\epsilon$. The master equation is
\beq
\frac{d}{dt} \rho_t(m) =J_t(m+1 ,m)- J_t(m-1 ,m)
\eeq
with $J_t(m ,m+1) = \rho_t(m) w_t(m,m+1)-\rho_t(m+1) w_t(m+1,m)$. Using the Taylor expansion  $\rho_t(m \pm 1)=\rho_t(m) \pm \partial_m \rho_t(m)$, we can rewrite the discrete currents as
\beq
J_t(m ,m \pm 1)=  w_t(m,m \pm 1)\rho_t(m)-w_t(m \pm 1,m)(\rho_t(m) \pm \partial_m \rho_t(m)).
\eeq
To establish a link between these discrete currents and the current arising in the corresponding Fokker-Planck equation, we introduce the notations
$u_t(m+1) = w_t(m,m+1) - w_t(m+1,m)$ and $2D_t(m+1)= w_t(m,m+1) + w_t(m+1,m)$,
so that the discrete currents are
\bea
J_t(m ,m + 1) &=& \rho_t(m) (u_t(m)+\partial_m u_t(m)) - w_t(m+1,m) \partial_m \rho_t(m)\\
J_t(m ,m-1) &=& - \rho_t(m) u_t(m) + w_t(m-1,m) \partial_m \rho_t(m)
\eea
In the continuous limit, $\epsilon \ra 0$
\beq
\rho_t(m) \sim \epsilon \rho_t(x), \quad \partial_m \sim \epsilon \partial_x, \quad u_t(m) \sim u_t(x)/\epsilon, \quad D_t(m) \sim D_t(x)/\epsilon^2,
\eeq
and the discrete currents can be related to the current $J_t(x)$ entering the Fokker-Planck equation:
\beq
J_t(m ,m \pm 1) \sim \pm u_t(x) \rho_t(x) \mp D_t \partial_x \rho_t(x)= \pm J_t(x).
\label{FP current}
\eeq

Let us recall the expression given in Eq.~\ref{nu} for the local current $\nu_{t'}(m)$:
\bea
\nu_{t'}(m) &=&  \frac{J_{t'}(m,m + 1) \partial_{h} w_{t'}^{h}(m,m + 1)|_{h \rightarrow 0}}{\rho_{t'}(m)} \\
            &+& \frac{J_{t'}(m,m - 1) \partial_{h} w_{t'}^{h}(m,m - 1)|_{h \rightarrow 0}}{\rho_{t'}(m)},
\eea

Using the expression of the rates given in Eq.~(\ref{bilan2}), we have $\partial_{h} w^{h}_{t'}(m,m \pm 1) =  (O(m \pm 1)-O(m))\beta/2  \sim \pm \beta \epsilon \partial_xO(x)/2$. Inserting this in the above equation and using Eq.~\ref{FP current}, one obtains the following expression of the local current $\nu_{t'}$ in the continuous limit
\beq
\nu_{t'}(x)= \frac{\beta J_{t'}(x)}{\rho_{t'}(x)} \partial_x O(x).
\eeq
Together with the first term in the response given in Eq.~\ref{new R1}, one recovers from this the  response function given in Ref.~\cite{Chetrite2009_vol80}, namely
\beq
R(t,t')= \beta \frac{d}{dt'} \left\langle A(x_t) O(x_{t'}) \right\rangle - \beta  \left\langle A(x_{t'}) \frac{J_{t'}(x_{t'})}{\rho_{t'}(x_{t'})} \partial_{x_{t'}} O(x_{t'})    \right\rangle.
\eeq

\section{Generalization to multidimensional linear Langevin dynamics}
\label{multidimensional}

We now extend the results obtained in section \ref{ex: Langevin} for Langevin dynamics with
one degree of freedom to many dimensions. To that end, we start with the following multidimensional linear Langevin equation:
\begin{equation}
\dot{x_t}=N_{t}x_t+h_{t}+\eta_{t}, \label{eq:lanlin}
\end{equation}
where $\eta_t$ is a white noise such that $\langle \eta_t \rangle=0$ and $\left\langle \eta_{t}\eta_{t'}\right\rangle =2T_{t}\Gamma_{t}\delta(t-t')$. We denote $T_t$ a time-dependent temperature, $\Gamma_t$ a positive and symmetric matrix, and $N_t$ an arbitrary matrix.

For any $t\geq s$, we introduce the following matrix
\beq
T(t,s)  =  \overleftarrow{\exp}\left(\int_{s}^{t}duN_{u}\right),
\eeq
where the exponential is to be understood as time-ordered.
This matrix satisfies the useful identity that
for any $t\geq s\geq u$, $T(t,s)T(s,u)=T(t,u)$.
We also introduce the matrix
\beq
D(t,s)  =  2\int_{s}^{t}du T(t,u)T_{u}\Gamma_{u}T^{\dagger}(t,u),
\eeq
where the dagger denotes the transpose of a matrix.

The solution of Eq.~\ref{eq:lanlin} is
\begin{equation}
x_{t}=T(t,0)x_{0}+\int_{0}^{t}dsT(t,s)(h_{s}+\eta_{s}).
\label{solution}
\end{equation}
We assume that the initial condition $x_0$ is distributed according to
a Gaussian, with a characteristic mean $m_{0}\equiv\left\langle x_{0}\right\rangle $, and an initial covariance matrix $V_{0}^{ij}=\left\langle \left(x_{0}-\left\langle x_{0}\right\rangle \right)^{i}\left(x_{0}-\left\langle x_{0}\right\rangle \right)^{j}\right\rangle $.
From Eq.~\ref{solution}, we deduce the mean
\begin{equation}
m_{t}=\langle x_t \rangle = T(t,0)m_{0}+\int_{0}^{t}dsT(t,s)h_{s},
\end{equation}
and the symmetric matrix of the covariance
\begin{equation}
V_{t}=\langle \left( x_t -m_t \right) \left( x_t - m_t \right) \rangle = T(t,0)V_{0}T^{\dagger}(t,0)+D(t,0).
\end{equation}
As in the 1D case, we note that in this case too, $V_t$ is independent of the perturbation $h_t$.
Since $x_t$ is a Gaussian random variable, its distribution is
\begin{equation}
P_{t}\left(x,\left[h_{t}\right]\right)=\left(\det\left(2\pi V_{t}\right)\right)^{-\frac{1}{2}}\exp\left(-\frac{1}{2}(x-m_{t})V_{t}^{-1}(x-m_{t})\right),
\end{equation}
and therefore
\begin{equation}
\pi_{t}(x,h_{t})=\left(\det\left(2\pi V_{t}\right)\right)^{-\frac{1}{2}}
\exp\left(-\frac{1}{2}(x-\widetilde{m}_{t})V_{t}^{-1}(x-\widetilde{m}_{t})\right),
\end{equation}
with
\begin{equation}
\widetilde{m}_{t}=T(t,0)m_{0}+\left(\int_{0}^{t}dsT(t,s)\right)h_{t}.
\end{equation}

Since
\begin{equation}
\partial_{h_{t}}\left(\ln\pi_{t}(x,h_{t})\right)=
\left(\int_{0}^{t}dsT^{\dagger}(t,s)\right)V_{t}^{-1}(x-\widetilde{m}_{t}),
\end{equation}
The functional $\Y$ of interest here, takes the form
\begin{equation}
\Y_{T}=-\int_{0}^{T}dt \dot{h}_{t} \left(\int_{0}^{t}dsT^{\dagger}(t,s)
\right)V_{t}^{-1}(x_{t}-\widetilde{m}_{t}).
\end{equation}
From this, we obtain the average
\begin{eqnarray}
\left\langle \Y_{T}\right\rangle  & = & -\int_{0}^{T}dt\dot{h}_{t} \left(\int_{0}^{t}dsT^{\dagger}(t,s)\right)V_{t}^{-1}(m_{t}-\widetilde{m}_{t}), \\
 & = & \int_{0}^{T}dt\dot{h}_{t} \left(\int_{0}^{t}dsT^{\dagger}(t,s)\right)V_{t}^{-1}\left(\int_{0}^{t} du T(t,u)\left(h_{t}-h_{u}\right)\right),
 \end{eqnarray}
so that
\begin{equation}
\Y_{T}-\left\langle \Y_{T} \right\rangle =-\int_{0}^{T}dt\dot{h}_{t} \left(\int_{0}^{t}dsT^{\dagger}(t,s)\right)V_{t}^{-1}(x_{t}-m_{t}).
\end{equation}
Through explicit calculation, one can verify that
\begin{eqnarray}
\left\langle \left(\Y_{T}-\left\langle \Y_{T}\right\rangle \right)\left(\Y_{T}-\left\langle \Y_{T}\right\rangle \right)\right\rangle  \\ \label{cov Y}
 =2\int_{0}^{T}dt\dot{h}_{t} \left(\int_{0}^{t}dsT^{\dagger}(t,s)\right)
 V_{t}^{-1}\int_{0}^{t}dt'T(t,t')\left(\int_{0}^{t'} ds T(t',s)\right)\dot{h}_{t'}. \nonumber
 \end{eqnarray}
Using an integration by parts, one can check that
\beq
\int_{0}^{t}dt'T(t,t') \left(\int_{0}^{t'} ds T(t',s)\right) \dot{h}_{t'}  =
 \int_{0}^{t}dt'T(t,t')\left(h_{t}-h_{t'}\right).
\eeq
Taken together, these two equations imply the relation
\begin{equation}
\left\langle \left(\Y_{T}-\left\langle \Y_{T}\right\rangle \right)\left(\Y_{T}-\left\langle Y_{T}\right\rangle \right)\right\rangle =2\left\langle \Y_{T}\right\rangle.
\end{equation}
Since $\Y_T$ is a Gaussian variable, the detailed fluctuation relation given in Eq.~\ref{DFR simple} follows from this for the general linear Langevin dynamics.

\section*{References}
\bibliographystyle{unsrt}
\bibliography{Ma_base_de_papier}

\end{document}